\documentclass[final,1p,times]{article}

\usepackage{graphicx}
\usepackage{url}
\usepackage[ruled, linesnumbered,noend]{algorithm2e}
\usepackage{amsmath,amssymb}
\usepackage{multirow}
\usepackage[usenames,dvipsnames,svgnames,table]{xcolor}
\usepackage{caption}
\usepackage{subcaption}
\usepackage{authblk}

\newcommand{\assign}{$\,$:=$\;$}

\newcommand{\pling}{{\tt plingeling}}
\newcommand{\azu}{{\tt AzuDICI}}

\usepackage{amssymb}

\begin{document}

  \title{Cache Performance Study of Portfolio-Based Parallel CDCL SAT
    Solvers}


  \author[*]{Roberto As\'in}
  \author[**]{Juan Olate}
  \author[**]{Leo Ferres}

  \affil[*]{\small Department of Computer Engineering, Faculty of Engineering, Universidad Cat\'olica de la Sant\'isima Concepci\'on, Chile\\rasin@ucsc.cl}
  \affil[**]{\small Department of Computer Science, Faculty of Engineering, Universidad de Concepci\'on, Chile\\
  \{juanolate,lferres\}@udec.cl}

\maketitle

  \begin{abstract}
    Parallel SAT solvers are becoming
    mainstream. Their performance has made them win the past two SAT
    competitions consecutively and are in the limelight of research
    and industry. The problem is that it is not known exactly what is
    needed to make them perform even better; that is, how to make them
    solve more problems in less time. Also, it is also not know how well
    they scale in massive multi-core environments which, predictably, is 
    the scenario of comming new hardware. In this paper we show that cache 
    contention is a main culprit of a slowing down in scalability, and provide
    empirical results that for some type of searches, physically sharing
    the clause Database between threads is beneficial.

  \end{abstract}

\section{Introduction}

In this paper, we study the effect of cache performance on the
scalability of parallel solvers of the satisfiability problem in
hierarchical-memory, symmetric multi-processing systems; systems with
more than one processor that share memory. We find this topic
important for three reasons:

First, in the last years, parallel SAT solvers (henceforth, pSATs)
have been performing at the top of the SAT
Competitions\footnote{\url{http://baldur.iti.uka.de/sat-race-2010/},
  \url{http://www.satcompetition.org/}} (in 2011, all three wall-clock
time winners of the competition were parallel solvers).  Also in 2011,
pSATs and sequential SAT solvers were grouped into a single
competition track, which shows the widespread interest in pSATs by
research and industry. This appeal stems in part from the inherently
interesting properties of parallel algorithms, but also because of the
need of the community to do better in new application domains and
handle even larger and more complex CNF formulas in shorter times,
taking advantage of modern hardware.

Second, instead of increasing clock performance, chip manufacturers
are investing heavily on multicore architectures to improve
performance and lower power consumption (AMD released the 8-core
Opteron 3260 EE in late 2011, and Intel did the same with the Xeon
E5-2650, and its low power version, the Xeon E5-2650L in early 2012).
As Herb Sutter put it, ``the free lunch is over''
\cite{FreeLunchIsOver}, and by this he meant that software in general
will not be getting any faster by simply relying on faster processor
clocks, but by relying on how software scales in multicore systems.

Finally, modern memory architectures are not flat
Processor$\leftrightarrow$RAM architectures, but a hierarchy of
fast-but-small to slower-but-large memories with latencies and
capacities varying from 0.5{\it ns} access to 32{\it Kb} memories in
the L1 (first level) cache, to tens of nanoseconds for megabyte-large
memories like the L3 cache (usually the last level, LL cache), to
100{\it ns} for gigabyte-sized access times to memory such as main
memory DDR ram. Hierarchical memory architectures have a strong impact
on the performance of sequential software (e.g., in a row-major
representation of a matrix, memory transfers may be in the order of
the input divided by the size of the cache line, while memory
transfers for column scanning are in the order of the square of the
input). This impact is equal or bigger in the case of parallel
processes since, for most architectures, the cores in the system share
some level of cache memory.

\begin{figure}[tp]
  \centering
  \includegraphics[scale=1,bb = 62 118 296 372,clip]{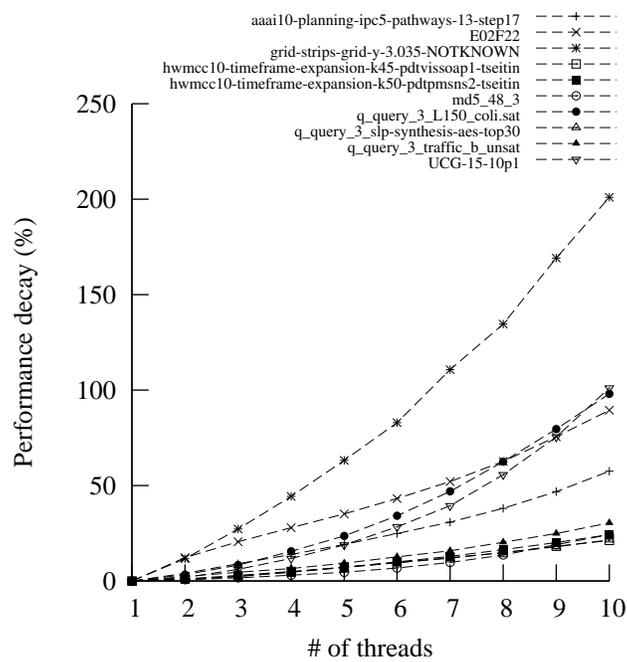}
  \caption{Performance decay of {\tt plingeling} when run over 10
    cores on a single processor using several standard benchmarks
    (their long names are in the legend.)}
  \label{fig:decay}
\end{figure}

Given the three reasons above, in this paper we are concerned with the
following questions: how do pSATs scale in hierarchical-memory
multicore architectures?  What is the effect of cache performance?
Our case study is the winner of the 2011 SAT Competition, {\tt
  plingeling}, a portfolio-based pSAT solver \cite{lingeling}. The
first experiment tested a \emph{modified} \pling\ on a 40-core machine
(four ten-core processors), varying the number of threads. This
instance of \pling\ was modified so that, for each {\em worker
  thread}, the {\em same search} would be performed (i.e. same
strategies, starting parameters and without lemma exchanging). This
allowed us to measure the impact of running $p$ threads on the same
physical CPU. In Figure \ref{fig:decay}, we see how the modified
\pling's performance decays sharply (around 30\% on average, up to
maximum of $\sim$200\%) when several solver instances are executed on
the same processor. This happens even though, in principle, instances
do not logically share any resources other than the common process
address space.  On the contrary, we would have expected that all
instances would perform similarly, plus or minus a small time
fraction. This is in fact what happens when \pling\ is run in four
different cores of four different chips, in different and even in the
same machine.

In order to find the reason of the performance decay, we have
developed a simple portfolio-based SAT solver (which we have called
\azu) that allows us to both replicate the behavior of \pling, and
then also experiment with alternative scenarios (such as physically
sharing information among threads) in order to take measures towards
the improvement of its cache performance. It is important to highlight
that our SAT solver does not compare to high-performance current
state-of-the-art solvers as is \pling\ itself. \azu\ serves as a
useful tool to test and analyze the behavior of
portfolio-based pSAT solvers.

The paper is structured as follows, highlighting our contributions in
the relevant sections: in Section \ref{sec:preliminaries} we introduce
the topic; that is, the general problem of SAT solving and
computational {\em sequential} SAT solving (Section \ref{sec:sesat})
and parallel SAT solving \ref{sec:parsat}. Section \ref{sec:cacheperf}
shows the cache performance of \pling, a state-of-the-art
portfolio-based SAT solver, when varying number of executing
threads. We show how cache contention significantly (and negatively)
impacts the performance of portfolio-based parallel SAT solvers. We
conclude here that pSAT solvers do not scale satisfactorily in shared-memory
parallel architectures. In Section \ref{sec:cperfphyscshare}, we
introduce \azu, a ``simple'' portfolio-based pSAT solver that
implements several levels of physical clause sharing. We also report
the results of several experiments that compare the cache performance
of these configurations, showing how physical clause sharing
significantly helps certain configurations (when all threads are
executing the same search) while either slightly or no help is
observed when pSAT solvers threads perform different searches.
Finally, Section \ref{sec:conclusions} closes the paper with some
discussion of the future of parallel SAT solving, situate our work in
the context of what has been done so far, and point the way for future
work.



\section{Preliminaries}
\label{sec:preliminaries}

\subsection{SAT and (Sequential) SAT solvers}
\label{sec:sesat}

Let $V$ be a fixed finite set of \emph{propositional variables}.  If
$v \in V$, then $v$ and $\lnot v$ are \emph{literals} of $V$.  The
\emph{negation} of a literal $l$, written $\lnot l$, denotes $\lnot v$
if $l$ is $v$, and $v$ if $l$ is $\lnot v$.  A \emph{clause} is a
disjunction of literals $l_1 \lor\ldots\lor l_n$.  A (CNF)
\emph{formula} is a conjunction of one or more clauses $C_1
\land\ldots\land C_n$.  A (partial truth) \emph{assignment} $M$ is a
set of literals such that $\{ v, \lnot v \} \subseteq M$ for no $v$. A
literal $l$ is \emph{true} in $M$ if $l \in M$, is \emph{false} in $M$
if $\lnot l \in M$, and is \emph{undefined} in $M$ otherwise.  A
clause $C$ is true in $M$ if at least one of its literals is true in
$M$.  It is false in $M$ if all its literals are false in $M$, and is
undefined in $M$ otherwise.  A formula $F$ is true in $M$, or
\emph{satisfied} by $M$, if all its clauses are true in $M$.  In that
case, $M$ is a \emph{model} of $F$.  If $F$ has no models then it is
\emph{unsatisfiable}.

The problem we are interested in is the \emph{SAT problem}: given a
formula $F$, to decide whether there exists a model of $F$ or not.
Since there exists a polynomial transformation
(see~\cite{Tseitin1968}) from any arbitrary formula to an
equisatisfiable CNF one, we will assume w.l.o.g. that $F$ is in CNF.

A program that solves this problem is called a \emph{SAT solver}. The
Conflict-Driven-Clause-Learning (CDCL) algorithm is nowadays at the
basis of most state-of-ther-art
SAT-solvers~\cite{gluclose,plingeling,cryptominisat}. This algorithm
has, at its roots, the very simple DPLL
algorithm~\cite{Davisetal1962CACM}. Thanks to work done mainly in
\cite{relsat,Chaff2001,GRASP1999IEEE,ZhangStickel1996AIMATH,EenSorensson2003SAT,picosat2008},
CDCL has evolved into an algorithm that allows modern SAT-solvers to
handle formulas of millions of variables and clauses. Algorithm
\ref{alg:cdcl} sketches the CDCL algorithm.

\begin{algorithm}[h]
  \SetAlgoLined \SetKwData{Conflict}{conflict} \SetKwData{Dl}{dl}
  \SetKwData{Lemma}{lemma} \SetKwData{Dec}{dec}
  \SetKwData{Status}{status} \SetKwData{Model}{model}
  \SetKwData{Result}{res}\SetKwData{BV}{BV}
  \SetKwInOut{Input}{Input}\SetKwInOut{Output}{Output} \Input{Formula
    $F=\{C_1,\ldots,C_m\}$} \Output{SAT OR UNSAT} \BlankLine \Status
  \assign UNDEF\; \Model \assign \{\}\; \Dl \assign $0$\;
  \While{\Status == UNDEF}{ (\Conflict, \Model) \assign BCP(\Model,
    $F$)\; \While{ \Conflict $\neq$ NULL}{ \If{\Dl == 0}{ \Return
        UNSAT\; } \Lemma \assign CONFLICT\_ANALYSIS(\Conflict, \Model,
      $F$)\;
      $F$ \assign F $\land$ \Lemma\; \Dl \assign
      LARGEST\_DL\_OF\_FALSE\_LITS(\Lemma, \Model)\; \Model \assign
      BACKJUMP\_TO\_DL(\Dl, \Model)\; (\Conflict, \Model) \assign
      BCP(\Model, $F$)\; }

    \If{\Status == UNDEF }{ \Dec \assign DECIDE(\Model, $F$)\; \Dl
      \assign \Dl +1\; \lIf{ \Dec = $0$ }{\Status \assign SAT}
      \Model \assign \Model $\cup$ \{ \Dec\}\; }

  } \Return \Status
  \caption{CDCL algorithm}
  \label{alg:cdcl}
\end{algorithm}

Basically, the CDCL algorithm is a backjumping search algorithm that
incrementally builds a partial assignment $M$ over iterations of the
\emph{DECIDE} and \emph{BCP} (Binary Constraint Propagation) procedures,
returning \emph{SAT} if $M$ becomes a model of $F$ or \emph{UNSAT} if no 
such model exists.

The DECIDE procedure corresponds to a branching step of the search and
applies when the unit propagation procedure cannot set any further
literal to true (see below). When no inference can be done about which
literals should be true in $M$, a literal $l^{dl}$ is ``guessed'' and
added to $M$ in order to continue the search.  Each time a new
decision literal is added to $M$ the\emph{ decision level} $dl$ of the
search is increased and we say that all literals in $M$ after $l^{dl}$
and before $l^{dl+1}$ belong to decision level $dl$.  For further
reading about the DECIDE procedure, we refer to
\cite{Chaff2001,EenSorensson2004,rsat}.

The BCP procedure applies when certain assignment $M$ falsifies all
literals but one, of an undefined clause $C$. So, if $l$ is the
undefined literal in $C$, in order for $M$ to become a model of $F$,
$l$ must be added to $M$ for $C$ to be satisfied. This procedure is
tested for every clause of the Formula and it ends when there is no
literal left to add to $M$ or when it finds a false clause.  In the
first case, BCP returns an updated model containing all such
propagations and the search continues.  In the second case it returns
the falsified clause, which we call a \emph{conflicting clause}.

Since BCP usually takes about 90\% of the total running time of a
typical modern SAT solver, many implementation techniques have been
proposed to make it more efficient. The algorithm that is implemented
in most current state-of-the-art SAT-solvers is known as the
\textit{two-watched literal scheme} \cite{Chaff2001}. The underlying
idea is that no clause with more than one literal will generate a
unit propagation or become conflicting if at least two of its literals
are undefined or, at least one of them is true. Hence, for each clause $C$, 
either $C$ is true or the
algorithm makes sure that two undefined literals exist. For this
purpose, two non-false literals are watched in every clause. For every
literal we keep a list of the clauses where it is being watched.  As soon as
some literal $l$ becomes false in the assignment, we visit every
clause $C$ in its watch list.  If the other watched
literal $l'$ of $C$ is true, then $C$ is satisfied and the invariant is
preserved. Otherwise we must find a non-false literal different from
$l$ and $l'$ to watch.  If we do not succeed and $l'$ is false, the
clause is conflicting; if we do not succeed and $l'$ is unassigned, we
unit propagate $l'$. For a detailed review on BCP and how different
implementations perform in hierarchical (cache) memory we refer to
\cite{ZhangMalik2003SAT}.

Since many problems have a great
percentage of binary clauses, many SAT solvers represent the set of
binary clauses as a \emph{graph of implications}. For each literal
$l$, a list of implied literals (literals that must be true
whenever $l$ is in $M$) is stored.  For example, the clause $l1 \lor
l2$ is stored by adding $l2$ to the implications list of $\lnot l1$
and so is $l1$ to the list of $\lnot l2$. Thus, BCP with binary
clauses becomes very efficient, since, to calculate the unit
propagation of a literal $l \in M$, it suffices to go through its
implications list and add every literal of the list to $M$ . In the
case that a given literal $l'$ of the list is false, then a
conflicting binary clause $\lnot l \lor l'$ is found and second
condition for BCP termination applies.

If BCP finds a conflicting clause, then two possibilities can
apply. If the decision level of the search is zero (i.e. no decisions have been
made) then the CDCL procedure returns UNSAT. If this is not the case,
a \emph{CONFLICT ANALYSIS} procedure is called. This procedure
analyzes the cause of such conflict (i.e. determines which decisions
have driven to the conflict) and returns a new clause (which we call a
\emph{lemma}) that is entailed by the original formula. Then, the
algorithm backjumps to an earlier decision level $dl'$ that
corresponds to the highest $dl'<dl$ among the false literals in the
lemma, and propagates with it.  CONFLICT ANALYSIS works in such a way
that, when backjumping and propagating with the lemma, the original
conflict is avoided. The lemmas learned at CONFLICT ANALYSIS time are
usually added to the formula in order to avoid similar conflicts and
can also be deleted (when the formula is too big and they are no
longer needed). For details of CONFLICT ANALYSIS, we refer to
\cite{GRASP1999IEEE,Zhang2001} and for lemma deletion heuristics to
\cite{Relsat97,Goldberg2002DATE,gluclose}. For a complete review of
this algorithm as well as proofs over its termination and soundness we
refer to \cite{Nieuwenhuisetal2006JACM}.

\subsection{Parallel SAT solvers}
\label{sec:parsat}

Parallel SAT solvers are not as mature as sequential ones and it is
still not clear which path to follow when designing and implementing
such new solvers. In this section we will briefly present the two main
approaches used in parallel SAT solvers for shared memory
architectures\footnote{The revision on SAT solving in distributed
  memory architectures is out of the scope of this paper.}.  We mainly
classify the parallel SAT solvers for shared memory architectures into
two categories: \emph{portfolio-approach} and {search-space splitting} solvers.

The main idea behind portfolio approach solvers is the fact that
different strategies/parameters of CDCL sequential solvers or even
different kinds of sequential solvers perform better for different
families of SAT problems.  In sequential CDCL SAT-solving, for
example, there exist several parameters/strategies related to the
algorithm's heuristics for restarting, deciding or cleaning the clause
database. Taking this into consideration, a portfolio approach is very
straightforward: run a group of sequential solvers in different
threads, each with different parameters and/or different
strategies. This idea can be easily extrapolated to other non-CDCL SAT
solvers.  The time the portfolio-based parallel solver will take to
solve the problem will be the time of the fastest thread in the group
of solvers running in parallel.  Differences between this kind of
solvers lie in whether the clause database should be physically
shared~\cite{Sartagnan} or, otherwise, if each thread should have its
own database. If this second approach is taken, it is possible to
implement the solver so that its different threads interchange lemmas
according to serveral policies: aggressively~\cite{ManySAT} or
selectively~\cite{plingeling} or avoiding communications between
threads at all~\cite{ppfolio}.

Search space splitting solvers do not run different solvers in
parallel, but run one solving instance that splits the search space
into disjoint subspaces. A common strategy to divide the search space
is to use \emph{guiding paths}~\cite{psato}. A guiding path is a
partial assignment $M$ in $F$, which restricts the search space of the
SAT problem. A solver that divides its search space with guiding paths
will assign threads to solve $F$ with the given $M$ of the guiding
path the thread was assigned to. Once a thread finishes searching a
guiding path with no success, it can request another to keep searching
(we refer to \cite{paMiraXT} for further explanations).

Both parallelization strategies (portfolio approach and
search-space-splitting approach) were and are currently being applied
to shared memory parallel computers (e.g. \cite{plingeling}) as well
as to distributed memory ones~(e.g. \cite{paMiraXT}). For a further
review on shared memory parallel SAT solving, we refer to
\cite{survey-psolvers}

In what follows, we focus on solvers implementing the portfolio
approach since this has reported the best results in recent papers and
competitions.

\subsection{Hardware and tools description}
\label{sec:march}

Since 2005, Intel has produced chips with more than one physical
processing core in order to speed up execution by sidestepping the
difficulty of producing chips with faster clocks. In these new
machines, each core has private small memories (called L1 cache, and
sometimes L2) and progressively bigger (but slower) shared memories
(usually L2 and L3). These effectively constitute a ``memory
hierarchy'' which needs to be kept coherent {\em to give the illusion
  of a single shared memory}. For instance, the architecture of two
machines used to run the tests were:
 
\begin{itemize}
\item Machine $K$: a dual-processor 6-core Intel Xeon CPU (E5645)
  running at 2.40GHz, with a total of 12 physical
  cores. Hyperthreading was disabled. The computer runs Linux
  3.0.0-15-server, in 64-bit mode.
\item Machine $I$: a quad-processor 10-core Intel Xeon CPU (E7-4860)
  running at 2.27GHz, for a total of 40 physical cores. Hyperthreading
  was enabled, but we never ran our code in more than 40 cores (Cores
  0-39) with the care that a process was always bind to an unassigned core. 
  Each core has one processor unit (PU), with separate L1d
  (32KB) and L2 (256KB) caches. They share a 30MB L3 cache. Main
  memory is 256GB. The computer runs Linux 2.6.18-194, in 64-bit mode.
\end{itemize}

We used two testing computers for technical reasons. Since we were not
the administrators of machine $I$, we did not have access to the tools
we needed for some of the experiments. On the other hand, Machine $I$
had more than three times the number of physical cores of Machine $K$,
providing stronger results. In any case, in those experiments where
results could be compared (those which measured relative time, for
example), we made the effort to compare them, and their behavior was
(in relative terms) always the same, rendering the results
generalizable.

\section{Cache performance without physical clause sharing}
\label{sec:cacheperf}

The \pling\ program is a portfolio-based pSAT solver that has won the
parallel-track of both the 2010 SAT
Race\footnote{\url{http://baldur.iti.uka.de/sat-race-2010/}} and the
2011 SAT
Competition\footnote{\url{http://satcompetition.org/}}. Briefly, when
\pling\ is called, it launches several \emph{worker} threads
(operating system threads, such as POSIX threads) that differ in their
random seeds, some heuristic values and the intensity of some formula
preprocessing methods. Each worker performs its individual search
separately and, whenever one of them finds a solution, it is reported
and the other workers are interrupted.  Regarding information sharing,
each of \pling's workers mantain its own clause Database and they only
exchange the Unitary Clauses they find during search.

In this Section, we report our experimentation with this
state-of-the-art solver and carefully analyze its performance in Cache
and the effect of such Cache-behaviour in the overall performance of
the solver.

\subsection{Modified \pling}
\label{sec:modifiedpling}

One of the advantages we assume of parallel computing is that the more
cores we add, the better performance we will obtain. This should be
also true for \pling, since the only difference of adding more threads
(assuming we have one thread per core) is that we will have a greater
variety of solver strategies trying to solve the same problem, and
also some logical clause sharing among threads. These are all valid
assumptions in theory, but our empirical results show that increasing
the number of threads also carries a considerable decrease in
performance for portfolio solvers like {\tt plingeling} due to cache
misses. In what follows, we go into detail.

Multicore shared memory systems have their cores sharing the same last
level cache (LLC) memory. The last level cache size in modern machines
has few megabytes and is usually not enough to hold all the data
required by a SAT instance. Therefore, there will be inevitably some
communication between the LLC and the main memory. The time cost of
communication between the CPU and the LLC cache is much lower than
between the CPU and the main memory, so we would like to keep data
transfers from main memory to a bare minimum.

Portfolio SAT solvers that only share clauses logically have to keep a
complete database of clauses for each thread's use. So as we add more
threads, the solver has greater needs of memory. But since usually all
cores (in the same chip) share the same LLC, all threads will have a
lower chance of finding their data in the LLC as we add more
threads. In this scenario, what we would expect to observe, as we have
in our experiments, is a considerable decrease in performance when
adding threads, simply because we incur in more LLC cache misses when
the amount of data to be manipulated by different threads
increases. We do not usually appreciate this negative performance
impact in these type of solvers, because different threads implement
different SAT solving strategies, so the solving time will mostly
depend on the fastest solving thread, shadowing the negative
performance impact of copying the clause database in each thread.

For experimentation purposes, we modifed \pling\ in such a way that
each thread did exactly the same search. For this, we initialized each
thread with the same random seed, heuristic values and clause
database.  Furthermore, in order to keep them searching in the same
way, we disabled clause sharing between threads (which in \pling\
corresponds to disabling the interchange of found units) and assure
that cleanup policies and algorithms were the same. In these
experiments, we would expect, theoretically, that adding more threads
would have no impact in the solving time, because all cores would be
exactly making the same search with their own data. However, in
practice, we found that the performance decay of having ten threads
spread over ten cores ranged from about 21\% (1.21 times slower) to
200\% (3 times slower) of the total time one thread would take (Figure
\ref{fig:decay}). Possible reasons for this behavior may be due to
several factors in modern SMP architectures. However, sharing
resources (such as the caches, communication and/or synchronization,
or main memory) could be seen as the main suspects. To find out, we
ran another experiment where a \pling\ instance performing the same
search with four threads was executed on different physical CPU {\em
  chips} (rather than cores), and, to compare, we ran the same
experiment on four cores of the same CPU chip.


\begin{figure}
\centering
\begin{subfigure}{.5\textwidth}
  \centering
  \includegraphics[width=.9\linewidth,bb=59 90 294 325,clip]{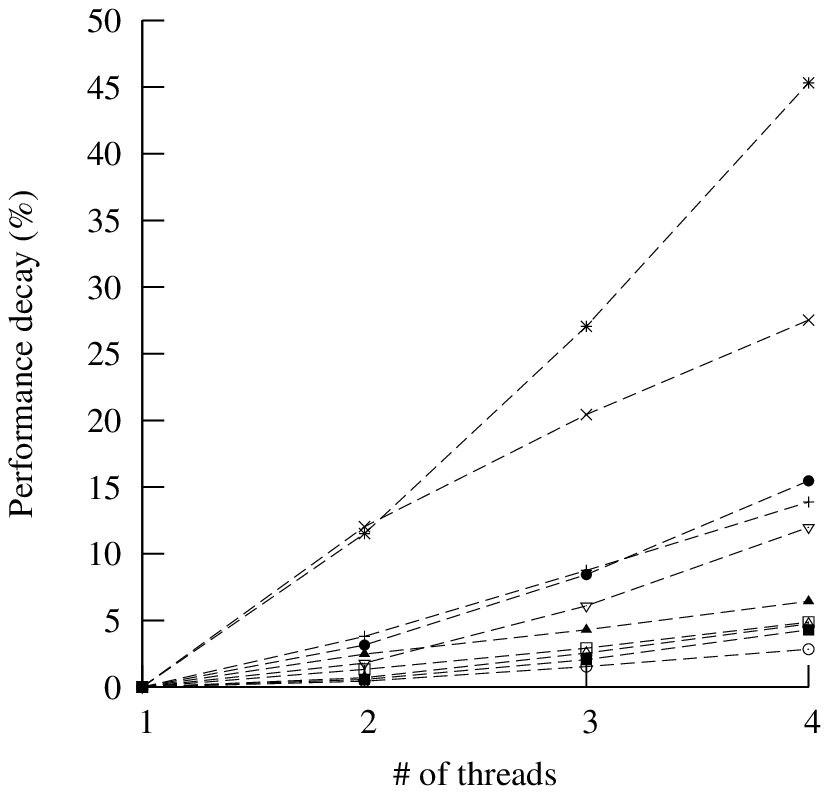}
  \caption{\pling\ same chip}
  \label{fig:4coressame}
\end{subfigure}%
\begin{subfigure}{.5\textwidth}
  \centering
  \includegraphics[width=.9\linewidth,bb=59 90 294 325, clip]{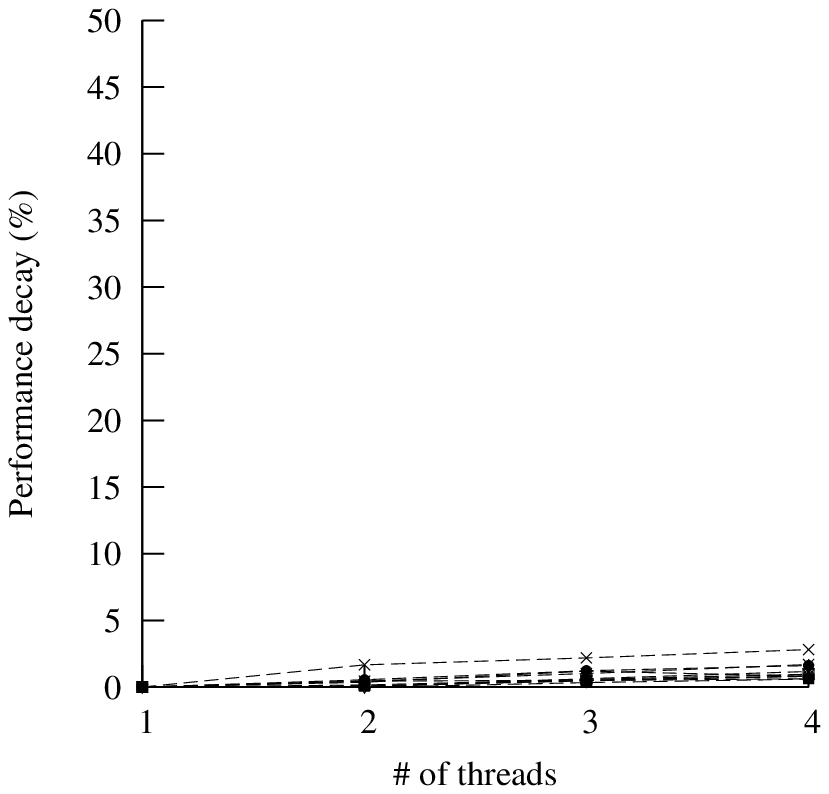}
  \caption{\pling\ different chip}
  \label{fig:4coresdiff}
\end{subfigure}
\caption{Modified \pling\ performance decay}
\end{figure}



\begin{figure}[h]
  \centering
  \includegraphics[bb = 62 118 296
  342,clip]{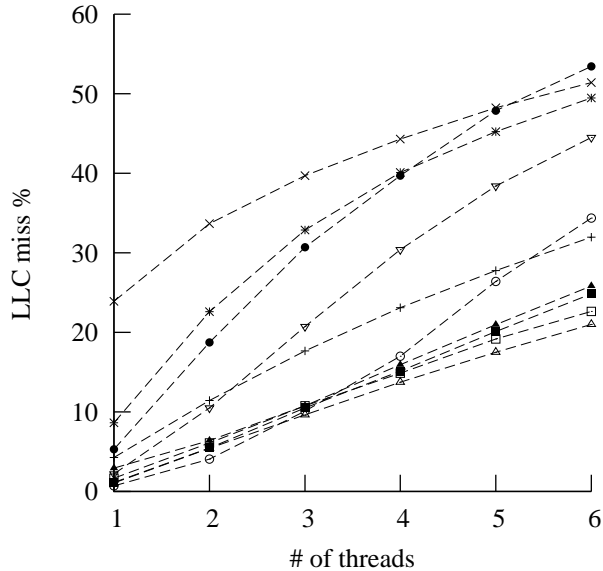}
  \caption{LLC statistics for modified \pling}
  \label{fig:LLCStats}
\end{figure}

As can be seen in Figure \ref{fig:4coresdiff}\footnote{This is Figure
  \ref{fig:decay}, showing only four cores}, executing the solver on
{\em different} CPU chips does not impact performance, while executing
it on the {\em same} CPU chip incurs in a significant performance
decay. According to the results above, the only shared resource that
could impact performance when run in one chip is the LLC or Last Level
Cache. To effectively measure the involvement of the LLC, we used the
{\tt perf} tool\footnote{\url{perf.wiki.kernel.org}. The {\tt perf}
  utility was the reason why we could not use machine I, it was only
  added in kernel version 2.6.31.}. The {\tt perf} tool is a hardware
abstraction over hardware counters of the different CPU chips
integrated in the Linux kernel to access profiling information on
retired instructions, branch misprediction, and in particular for our
purposes counting percentage of cache misses over cache hits ({\em
  LLC-load-misses}). As Figure \ref{fig:LLCStats} strongly suggests,
the performance decay observable in Figure \ref{fig:decay} is due to
several solving threads {\em thrashing} the cache and thus wasting
much more time retrieving their individual data from main memory than
their single-threaded counterparts.

\subsection{\pling\ scalability}
\label{sec:scale}

In this section, we provide an overview of how \pling\ behaves at a
larger scale. So far, thanks to the experiment above, we know that the
more threads we add, the more cache hits/misses impacts negatively on
performance. However, we also know that adding threads also adds new
(and possibly successful) strategies. This results in a trade-off
between cache contention versus portfolio-approach benefits. To find
out where the trade-off equilibrium lies, we ran the original \pling\
over 208 standard benchmarks taken from past SAT Races and
Competitions (see link above), varying the number of threads from one
to ten on a single chip with 10 physical cores. These 208 benchmarks
are the newly reported industrial/application benchmarks of
competitions: the 2009 and 2011 SAT Competition and SAT Race 2010.
\begin{table}[hhh]
  \centering
  \begin{tabular}[h]{ccc}
    \hline
    {\bf Threads} & {\bf \# Problems solved} & {\bf Total time} \\
    \hline
    1             & 113                      & 101399           \\
    2             & 121                      & ~95745           \\
    3             & 119                      & ~93854           \\
    4             & 122                      & ~90412           \\
    5             & 124                      & ~87953           \\
    6             & 124                      & ~89506           \\
    7             & 127                      & ~87416           \\
    8             & 124                      & ~88434           \\
    9             & 124                      & ~88931           \\
    10            & 125                      & ~89003           \\
    4 in 4 CPUs   & 126                      & ~88092           \\
    10 in 4 CPUs\footnote{Divided into 3 processes for CPUs 1 and 2, and two processes for CPUs 3 and 4.}  & 129                      & 85224            \\
    40 in 4 CPUs  & 123                      & 92387            \\
    \hline
    \hline
  \end{tabular}    
  \caption{Scalability}
  \label{tab:scal}
\end{table}

\begin{figure}[htp]
  \centering
  \includegraphics[bb=62 116 301 340, clip]{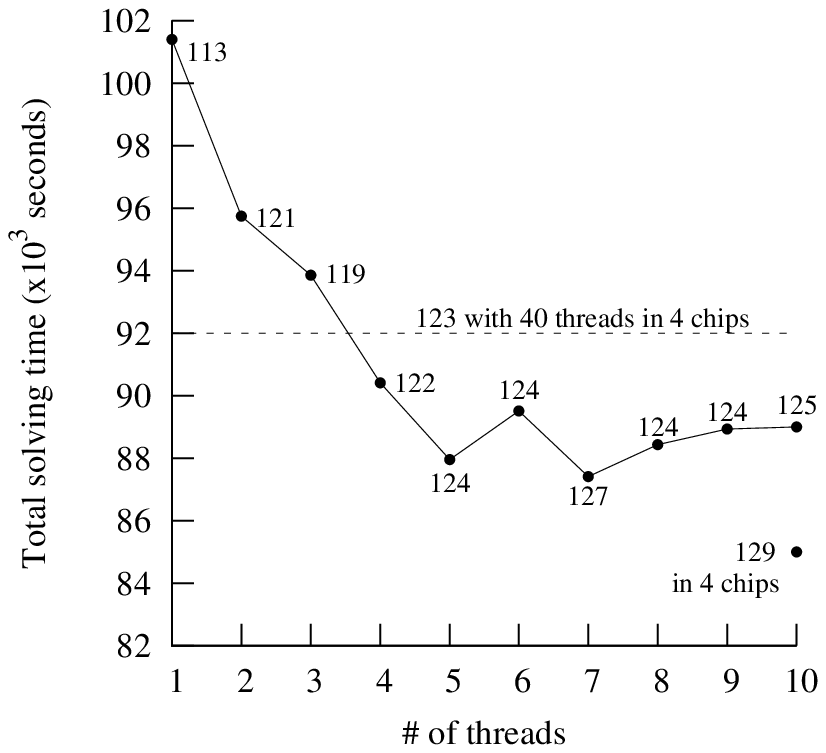}
  \caption{Total solving time per number of threads.}
  \label{fig:pperfpthread}
\end{figure}

Figure \ref{fig:pperfpthread} and Table \ref{tab:scal} show that up
until the fifth thread, scalability is good, but from then on, the
number of solved problems and total time reaches a plateau. This means
that \pling\ cannot scale up on the number of cores sharing an LLC. It
is important to notice that executing the same ten-thread solver in
four different physical CPU chips solves more problems in less time,
while executing a 40-thread solver (10 threads per chip) behaves worse
than the four-threaded single-chip version. This effectively means
that sharing cache among threads has a negative impact in the overall
behavior of modern portfolio-approach-based parallel SAT solvers.

\section{Cache Performance with physical clause sharing}
\label{sec:cperfphyscshare}

\subsection{\azu}
\label{sec:azudici}

It is clear that the problem with cache-misses is tightly related to
the way in which BCP is implemented \cite{ZhangMalik2003SAT} and,
therefore, with the way in which clauses and watches are programmed.
Since \pling\ keeps a separate clause database for each thread, it is
possible that sharing data could improve cache performance, because we
would have a smaller amount of total data to propagate with.  Based on
other parallel SAT-solvers that share their clause database, mainly
{\tt SArTagnan} and {\tt MiraXT}, we decided to implement \azu, a
basic CDCL SAT-solver with the purpose of improving the BCP
performance in portfolio-based pSAT solvers. Three versions of \azu\
were implemented, one in which each thread keeps a separate database
clause; another in which, as in MiraXT, shares all clauses physically;
and a hybrid one that only shares the binary implication lists.

\subsubsection{The general structure of \azu}

\azu\footnote{You can find the latest implementation of \azu\ at
  \url{https://github.com/leoferres/azu}.} is a standard CDCL solver
based on \pling, {\tt barcelogic} and {\tt miraXT}.  In particular,
\azu\ implements binary implication lists for the propagation with
binary clauses, and the two-watched\cite{ZhangMalik2003SAT} for BCP
with clauses of more than two
literals. \azu\ also implements the 1-UIP algorithm\cite{GRASP1999IEEE,Chaff2001} for conflict
analysis, the lemma simplification
algorithm used in {\tt PicoSAT}, Luby restarts \cite{LubySZ93}, a
policy for lemma cleaning that keeps only binary and ternary lemmas,
and more than four-literal lemmas that have participated in a conflict
since the last cleanup. Finally, \azu\ also incorporates the EVSIDS
heuristic for branching literal decisions \cite{EenSorensson2004}.

\subsubsection{Shared-none}

This version works as \pling, it does not share any clause physically.
Each thread keeps its own independent database of clauses and
propagates with it. Note that the fact that we are not sharing data
physically does not mean threads cannot share information. They could,
for example, share unit clauses through message passing between
threads, just as \pling\ does. We are interested in measuring the
impact of sharing data physically on cache performance, and not the
benefits for the search of sharing information itself.

\subsubsection{Shared-bins}

\begin{figure}[tp]
  \centering
  \includegraphics[scale=1.0]{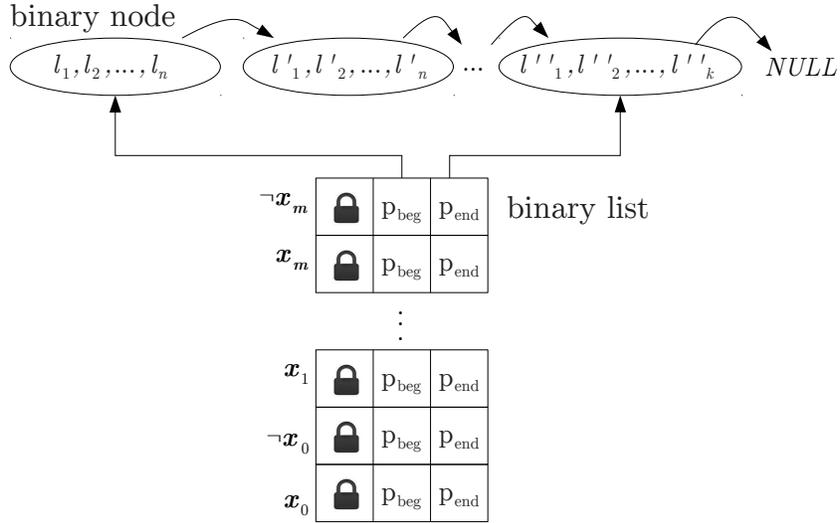}
  \caption{Binary clause database}
  \label{fig:shared bins}
\end{figure}

The Shared-bins version shares the binary implication lists. All
threads have access to the same physical data, they all can modify and
read this structure. Figure \ref{fig:shared bins} is a schematization
of our binary implication lists structure. We have an array of
\textit{binary lists}, one for each literal. A binary list is
basically two pointers, one to a first \textit{binary node} and
another to the last binary node associated with that list. A binary
node is an array of literals that also has a pointer to another binary
node. The amount of literals a binary node can hold will depend on the
size of the cache line we are working with; it will have as many
literals as a cache line can hold. The literals implied by the literal
associated to a binary list will be the ones in the binary node
referenced by that binary list pointer and the subsequently referenced
binary nodes.

When a thread wants to add the clause $\{l_i,l_j\}$, it must look for
the binary list associated with $\neg l_i$ and go to the last node
linked to that binary list. If there is enough space in that node to
add another literal, then it adds $l_j$. If the node is full, then it
must create a new node with the $l_j$ literal, insert it at the end of
the linked list of nodes and update the binary list last node pointer.
It does the same for the binary list of $\neg l_j$.

To ensure consistency of data when multiple threads are inserting,
each binary list has a lock.  If a thread is inserting a new
implicated literal, it first locks the binary list where it is
inserting and then proceeds to insert. If by chance another thread
wants to insert in the same binary list, it must wait till the lock is
freed. Since adding binary clauses is not frequent, and the event that
it would happen in the same binary list is even less frequent, the
contention that these locks generate is unnoticeable in our
experimental results.

\subsubsection{Shared-all}

\begin{figure}[tp]
  \centering
  \includegraphics[scale=0.6]{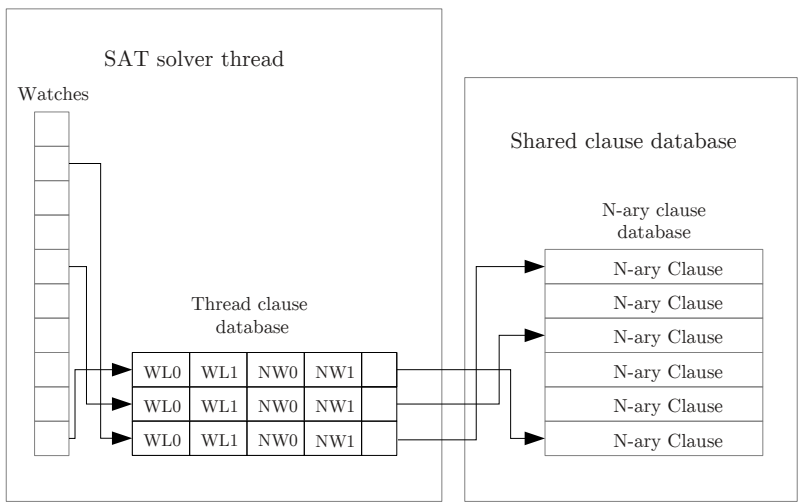}
  \caption{The thread clause database and n-ary clause database}
  \label{fig:azu design}
\end{figure}

In the shared-bins solver we had no need to modify the usual two
watched literal scheme used in propagation. This is not the case for
this version that shares n-ary clauses physically. For the
implementation of the two watched literal scheme, it is necessary to
keep track of which literals are being watched in each n-clause. For
instance, in a sequential solver, a typical implementation would
consist in watching the first two literals of the clause. In portfolio-based
pSAT solvers that physically share de clauses, since
several threads could be accessing the same clause, changes to
the clause are not feasible. It is impossible since threads of
the portfolio may be watching different literals of a same
clause. Instead, we have used a similar approach to that used in {\tt
  MiraXT}, where each thread keeps track of the literals being watched
in each clause. Figure \ref{fig:azu design} is a schematization of how
each thread worker relates with the n-ary clause database. Each SAT
solver thread has a vector of pointers to thread clauses called
\textit{watches}, and each literal present in the SAT problem has a
position associated with this watches vector. A thread clause has two
watched literals (WL0 and WL1), two pointers to another thread clause
(NW0 and NW1) and a pointer to an actual n-ary clause in the n-ary
clause database. W0 and W1 keep track of the literals being watched by
the thread for a given n-ary clause. NW0 and NW1 point to the next
thread clauses where WL0 and WL1 are also being watched. The n-clause
also has a flag for each worker thread to identify which ones are
using that clause for propagation.

To insert a new n-ary clause, we first make sure that the clause does
not exist in the database. If it does not exist, we create the
n-clause, set the current thread flag to true and add it to the
database.  On the other hand, if it does exist, we just toggle the
corresponding thread flag of the n-clause to true. The ``insert''
procedure is locked so that two different threads can not insert at
the same time. In our experiments we have not noticed any considerable
overhead caused by this lock. In fact, after profiling, the time spent
by the ``insert'' function is negligible in the total running time of
the program, with or without locks.

To find out whether sharing the clause database was beneficial to the
same-chip portfolio-based solver, we ran \azu\ in three different
versions (sharing all the clauses, sharing none of the clauses and
sharing only binary clauses) on eight problems using one to six
threads. Each run was repeated five times to mitigate potential system
noise.


In the following subsections we present the results of two experiments
measuring cache misses. In section \ref{sec:samesearch} we ran the
three different versions of \azu\ where each thread executed the same
search, which resembles the experiments done using modified \pling
(see section \ref{sec:modifiedpling}). In section
\ref{sec:diffsearch}, the three versions of \azu\ were used to test the
canonical work of SAT solvers, where each thread executed a different
search.

\subsection{Same search experiments}
\label{sec:samesearch}

For this experiment, \azu\ was modified to carry out the {\em same}
search in each thread (i.e., there is no lemma sharing among
threads). For each \azu\ version, we measured the time needed to solve
each benchmark, and the percentage of LLC misses. The results for this
are shown in Table \ref{tab:sstime} and below.

Notice that the datasets we chose for this experiment were influenced
by the early state of development of \azu. Our solver is not
optimized, and has been implemented for the purposes of
experimentation. Thus, the datasets are generally ``easier'' so that
\azu\ can solve them. Contrariwise, we do not use the benchmarks we
used for \azu\ for \pling, because these are solved so fast that
scalability cannot be reliably measured. In other words, we have
divided the whole dataset of problems into ``easy'' and ``hard''
problems. We have operationalized ``easy'' problems as those that
\azu\ can solve in the span of five to fifteen minutes, while \pling\
takes less than one minute. ``Hard'' problems, in turn, are those that
\pling\ takes between five and fifteen minutes to solve. We use
hard problems in \azu\ for different search experiments with a timeout 
of 15 minuts and we don't use easy
problems in \pling, since, for the latter, results would be tainted by
system noise. Besides, given the nature of CDCL solvers, the
size of the clause database will be increasing with execution time,
and this size increase is where cache contention manifests itself more
evidently.

\begin{table}
  \scriptsize
  \centering
  \begin{tabular}[h]{rccccc} \hline
    Dataset                    & T2   & T3   & T4   & T5   & T6   \\ \hline
    \multirow{3}{*}{manol-pipe-c10b}      & 1.16 & 1.33 & 1.46 & 1.58 & 1.66 \\
                               & 1.14 & 1.28 & 1.40 & 1.50 & 1.58 \\
                               & 1.11 & 1.24 & 1.33 & 1.44 & 1.53 \\ \hline
    \multirow{3}{*}{manol-pipe-c6bid\_i}  & 1.17 & 1.32 & 1.43 & 1.54 & 1.61 \\
                               & 1.15 & 1.29 & 1.39 & 1.48 & 1.55 \\
                               & 1.12 & 1.23 & 1.33 & 1.41 & 1.53 \\ \hline
    \multirow{3}{*}{manol-pipe-c6nidw\_i} & 1.18 & 1.33 & 1.44 & 1.54 & 1.61 \\
                               & 1.16 & 1.30 & 1.39 & 1.48 & 1.55 \\
                               & 1.12 & 1.22 & 1.32 & 1.41 & 1.52 \\ \hline
    \multirow{3}{*}{manol-pipe-c7idw}     & 1.19 & 1.33 & 1.42 & 1.50 & 1.56 \\
                               & 1.15 & 1.27 & 1.36 & 1.41 & 1.48 \\
                               & 1.15 & 1.24 & 1.31 & 1.38 & 1.47 \\ \hline
    \multirow{3}{*}{manol-pipe-cha05-113}    & 1.14 & 1.28 & 1.41 & 1.53 & 1.61 \\
                               & 1.12 & 1.24 & 1.35 & 1.45 & 1.53 \\
                               & 1.11 & 1.22 & 1.29 & 1.38 & 1.47 \\ \hline
    \multirow{3}{*}{anbul-dated-5-15-u}    & 1.26 & 1.49 & 1.67 & 1.81 & 1.91 \\
                               & 1.24 & 1.45 & 1.60 & 1.73 & 1.81 \\
                               & 1.15 & 1.39 & 1.50 & 1.64 & 1.74 \\ \hline
    \multirow{3}{*}{ibm-2002-31\_1r3-k30}   & 1.10 & 1.22 & 1.29 & 1.37 & 1.42 \\
                               & 1.10 & 1.21 & 1.28 & 1.35 & 1.39 \\
                               & 1.09 & 1.16 & 1.23 & 1.30 & 1.34 \\ \hline
    \multirow{3}{*}{post-c32s-gcdm16-22}     & 1.09 & 1.17 & 1.23 & 1.30 & 1.33 \\
                               & 1.08 & 1.17 & 1.22 & 1.28 & 1.30 \\
                               & 1.06 & 1.15 & 1.20 & 1.26 & 1.29 \\ \hline
  \end{tabular}
  \caption{ Performance decay in percentage over the T1 running time
    for \azu\ version and number of threads (T). The first row of each
    dataset corresponds to Shared-None version, second row to
    Shared-Bin and third row to Shared-All.}
  \label{tab:sstime}
\end{table}

\begin{figure}
  \centering
  \includegraphics[scale=0.9,bb=62 121 299 340,clip]{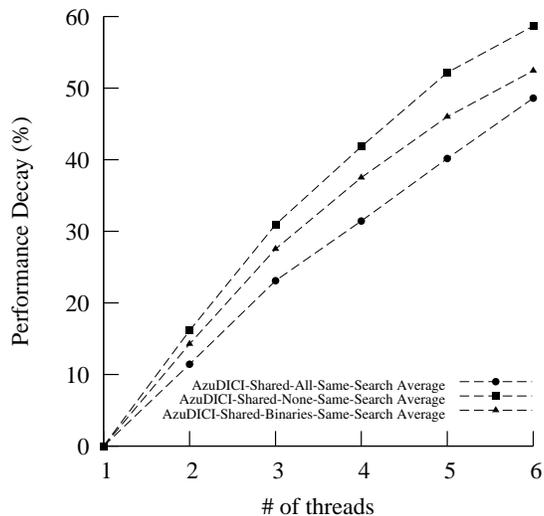}
  \caption{Average running time for AZUDici version and number of
    threads}
  \label{fig:ssruntimes}
\end{figure}

Table \ref{tab:sstime} and Figure \ref{fig:ssruntimes} show
performance decay over the one thread (T1=1) setting. Even if
performance is worse as we add more threads, shared-all will perform
consistently better than shared-binary, which will in turn perform
better (perhaps less noticeably) than shared-none. This is due to the
different levels of non-replication and physical sharing of the
database clause.

\begin{figure}[htp]
  \centering
  \includegraphics[scale=.9,bb=62 91 299 340,clip]{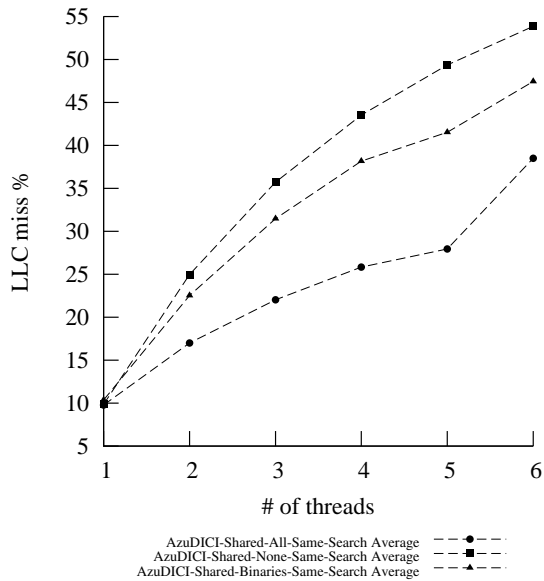}
  \caption{Average cache misses for AZUDici version and number of
    threads}
  \label{fig:sscachemisses}
\end{figure}

\begin{table}[h]
  \scriptsize
  \centering
  \begin{tabular}[h]{rcccccc} \hline
    Dataset                    & T1        & T2        & T3        & T4        & T5        & T6        \\ \hline
    \multirow{3}{*}{manol-pipe-c10b}      & 2.6(2.4)  & 15.7(0.6) & 27.2(0.1) & 36.4(0.2) & 43.2(0.1) & 48.8(0.0) \\ 
                               & 3.5(5.0)  & 13.5(2.9) & 21.0(3.6) & 27.5(2.9) & 31.2(1.7) & 38.6(0.8) \\ 
                               & 3.4(4.8)  & 9.6(8.2)  & 15.4(5.1) & 17.7(5.3) & 21.2(4.4) & 29.2(2.5) \\ \hline
    \multirow{3}{*}{manol-pipe-c6bid\_i}  & 11.8(0.3) & 27.2(0.1) & 37.5(0.1) & 44.8(0.1) & 50.3(0.0) & 54.7(0.0) \\ 
                               & 12.5(0.2) & 24.0(2.6) & 32.6(2.0) & 38.6(1.2) & 43.3(2.7) & 49.0(0.5) \\ 
                               & 12.1(0.4) & 18.0(7.1) & 21.1(0.1) & 25.7(3.2) & 28.3(1.0) & 43.3(1.3) \\ \hline
    \multirow{3}{*}{manol-pipe-c6nidw\_i} & 13.2(0.2) & 29.0(0.1) & 39.3(0.1) & 46.5(0.0) & 52.1(0.1) & 56.3(0.0) \\ 
                               & 14.0(0.6) & 26.2(2.3) & 35.4(1.4) & 41.2(2.4) & 44.1(0.6) & 50.3(0.5) \\ 
                               & 13.5(0.3) & 19.5(6.4) & 22.6(0.1) & 26.4(0.1) & 29.5(0.1) & 43.7(1.8) \\ \hline
    \multirow{3}{*}{manol-pipe-c7idw}     & 13.7(0.6) & 30.1(0.1) & 38.7(0.1) & 44.6(0.1) & 49.0(0.0) & 52.6(0.1) \\ 
                               & 14.8(0.4) & 24.6(3.7) & 30.8(3.3) & 35.8(2.5) & 34.4(1.6) & 42.0(2.0) \\ 
                               & 15.3(0.4) & 22.4(6.2) & 22.9(0.1) & 25.2(0.1) & 27.2(0.1) & 36.4(4.5) \\ \hline                             
    \multirow{3}{*}{manol-pipe-cha05-113}    & 1.8(2.3)  & 13.2(1.1) & 23.2(0.6) & 31.8(0.3) & 38.6(0.1) & 44.4(0.0) \\ 
                               & 2.4(4.2)  & 10.6(1.4) & 17.4(1.8) & 23.7(1.0) & 27.1(1.2) & 34.6(0.5) \\
                               & 2.5(6.0)  & 8.5(0.8)  & 13.0(8.2) & 15.3(5.9) & 16.8(2.4) & 24.2(2.8) \\ \hline  
    \multirow{3}{*}{anbul-dated-5-15-u}    & 8.6(0.6)  & 28.4(0.2) & 42.9(0.1) & 52.3(0.1) & 58.9(0.0) & 63.2(0.0) \\ 
                               & 8.9(0.6)  & 27.7(0.1) & 41.1(0.3) & 50.2(0.1) & 55.8(0.2) & 61.0(0.1) \\ 
                               & 6.5(9.1)  & 15.0(9.1) & 28.9(7.9) & 33.9(3.8) & 37.9(2.0) & 49.2(1.7) \\ \hline           
    \multirow{3}{*}{ibm-2002-31\_1r3-k30}   & 13.2(1.2) & 28.7(0.2) & 39.9(0.0) & 48.0(0.0) & 53.7(0.1) & 57.8(0.0) \\ 
                               & 13.0(0.3) & 28.2(0.2) & 39.0(0.1) & 47.1(0.0) & 52.4(0.3) & 56.3(0.2) \\ 
                               & 11.8(2.5) & 23.3(0.8) & 25.1(4.1) & 31.7(4.8) & 31.9(3.1) & 42.5(2.1) \\ \hline
    \multirow{3}{*}{post-c32s-gcdm16-22}     & 13.8(0.7) & 27.4(0.2) & 37.1(0.0) & 43.9(0.1) & 49.0(0.1) & 53.0(0.0) \\ 
                               & 13.8(0.1) & 25.4(2.7) & 34.4(0.8) & 41.2(0.3) & 44.0(1.0) & 47.6(0.9) \\ 
                               & 13.2(0.3) & 19.8(9.0) & 27.3(3.2) & 31.0(1.8) & 30.7(3.6) & 39.5(1.5) \\ \hline
  \end{tabular}
  \normalsize
  \caption{LLC misses (\%) for \azu\ version and number of threads (T). The first row of each dataset corresponds to Shared-None version, second row to Shared-Bin and third row to Shared-All  . Numbers in parentheses are standard deviations in \%.}
  \label{tab:sscachemisses}
\end{table}

On aggregate, comparing the performance of \azu\ as shown in Figure
\ref{fig:ssruntimes} and \ref{fig:sscachemisses}, it is evident that
the solver performs best in the shared-all setting, followed by the
shared-binary and finally by the shared-none setting in both running
time and cache misses. Thus, our implementation shows that physically
sharing the clause database is beneficial to avoid cache contention in
the rather artificial case of same search.

\subsection{Different search experiments}
\label{sec:diffsearch}

Although the previous experiments point out that physically sharing
the clause database between threads may lead to improve the cache
performance of the solvers, the results are not generalizable to a
full-featured SAT solver. It may be the case that while threads are
carrying out the same search, it is more likely that they will access
the same data. Whereas different search threads are clearly not
necessarily accessing the same data at the same time.

To find out how real portfolio-based SAT solvers implementing
different levels of physical clause sharing behave, we ran \azu\ in the
same three different versions as before (sharing-all, share-none and
share-bin) on the eight problems introduced in Section
\ref{sec:modifiedpling} using one to six threads and a 5-minute
timeout (we therefore did not include a running time table and
graph). The timeout characteristics was due to the fact that different
searches among threads may result in different (potentially better)
strategies, affecting execution time and rendering search behavior
effectively incomparable. Each run was repeated five times to clean up
potential system noise.


\begin{table}[htbp]
  \scriptsize
  \centering
  \begin{tabular}[h]{rcccccc} \hline
    Dataset                                                 & T1        & T2        & T3        & T4        & T5        & T6        \\ \hline
    \multirow{3}{*}{aaai10-\ldots-step17}                   & 3.0(5.2)  & 7.5(0.6)  & 14.0(0.3) & 21.5(0.3) & 27.3(0.2) & 34.1(0.1) \\
                                                            & 2.9(1.4)  & 6.4(3.8)  & 13.9(2.7) & 17.8(3.0) & 25.5(1.1) & 31.2(1.5) \\
                                                            & 2.8(0.9)  & 5.8(2.0)  & 12.3(1.8) & 16.5(1.4) & 22.6(0.2) & 29.0(1.3) \\ \hline
    \multirow{3}{*}{E02F22}                                 & 31.8(0.4) & 44.0(0.1) & 48.5(0.0) & 53.8(0.0) & 58.0(0.0) & 62.6(0.1) \\
                                                            & 26.8(0.2) & 40.9(5.1) & 47.5(3.4) & 54.7(1.5) & 58.9(2.3) & 62.3(1.9) \\
                                                            & 30.9(0.1) & 40.2(2.7) & 48.7(3.1) & 52.8(1.9) & 57.3(1.6) & 60.4(2.2) \\ \hline
    \multirow{3}{*}{grid-\ldots-3.035-NOTKNOWN}             & 0.2(3.5)  & 0.5(1.4)  & 0.5(0.9)  & 0.7(2.1)  & 0.8(2.9)  & 1.0(7.0)  \\
                                                            & 0.2(8.5)  & 0.7(4.3)  & 0.9(1.4)  & 1.0(3.9)  & 1.1(3.4)  & 1.4(2.2)  \\
                                                            & 0.3(1.1)  & 0.3(1.4)  & 0.6(4.2)  & 0.9(4.4)  & 1.1(4.0)  & 1.4(7.7)  \\ \hline
    \multirow{3}{*}{hwmcc10-\ldots-k45-pdtvissoap1-tseitin} & 22.0(0.4) & 33.6(0.1) & 44.2(0.1) & 50.6(0.1) & 54.9(0.0) & 59.5(0.0) \\
                                                            & 25.3(0.6) & 34.5(0.4) & 42.1(0.5) & 48.3(0.2) & 53.3(0.1) & 56.8(0.3) \\
                                                            & 25.1(0.1) & 34.1(0.6) & 41.8(0.3) & 48.3(0.3) & 52.6(0.2) & 56.0(0.3) \\ \hline
    \multirow{3}{*}{hwmcc10-\ldots-k50-pdtpmsns2-tseitin}   & 12.8(0.3) & 27.0(0.0) & 35.9(0.1) & 44.8(0.0) & 49.3(0.1) & 54.0(0.0) \\
                                                            & 14.0(0.6) & 26.0(0.9) & 35.8(0.7) & 40.9(1.3) & 46.7(0.2) & 49.8(0.5) \\
                                                            & 15.0(0.3) & 26.8(1.3) & 36.2(0.3) & 41.9(0.1) & 46.6(0.4) & 50.6(0.3) \\ \hline
    \multirow{3}{*}{md5\_48\_3}                             & 12.1(0.5) & 23.3(0.1) & 31.9(0.2) & 39.1(0.1) & 43.6(0.1) & 48.2(0.0) \\
                                                            & 13.4(0.3) & 23.0(2.8) & 32.0(0.2) & 37.0(0.2) & 41.8(2.1) & 46.1(0.1) \\
                                                            & 14.3(0.2) & 23.5(0.3) & 32.2(0.3) & 37.0(0.1) & 41.8(0.2) & 45.7(0.2) \\ \hline
    \multirow{3}{*}{q\_query\_3\_L150\_coli.sat}            & 23.1(0.8) & 39.0(0.2) & 50.2(0.0) & 58.2(0.1) & 62.1(0.0) & 66.0(0.0) \\
                                                            & 35.8(0.1) & 43.0(1.2) & 52.8(1.5) & 57.9(0.3) & 62.3(0.8) & 66.8(0.3) \\
                                                            & 34.5(0.3) & 36.6(2.8) & 50.4(1.3) & 56.0(0.5) & 61.2(0.7) & 63.6(0.5) \\ \hline
    \multirow{3}{*}{slp-synthesis-aes-top30}                & 11.5(0.7) & 20.4(0.2) & 29.3(0.1) & 36.8(0.1) & 42.3(0.1) & 47.7(0.1) \\
                                                            & 12.2(0.2) & 17.8(0.8) & 26.2(0.5) & 32.1(0.2) & 37.8(0.4) & 42.9(0.3) \\
                                                            & 13.2(0.2) & 18.3(0.2) & 26.3(0.2) & 31.7(0.4) & 37.4(0.3) & 42.1(0.2) \\ \hline
    \multirow{3}{*}{traffic\_b\_unsat}                      & 4.4(3.4)  & 17.4(0.7) & 31.9(0.3) & 44.3(0.1) & 52.1(0.0) & 59.8(0.0) \\
                                                            & 4.3(1.8)  & 16.7(1.2) & 31.4(0.2) & 41.7(0.2) & 51.4(0.1) & 58.7(0.0) \\
                                                            & 4.0(3.5)  & 14.6(0.3) & 28.4(0.5) & 37.5(0.4) & 46.3(0.2) & 53.8(0.1) \\ \hline
    \multirow{3}{*}{UCG-15-10p1}                            & 21.4(0.6) & 31.8(0.1) & 40.6(0.1) & 43.6(0.1) & 46.9(0.1) & 49.8(0.1) \\
                                                            & 20.7(0.7) & 31.7(3.0) & 40.9(2.4) & 46.2(2.0) & 53.5(0.8) & 55.5(0.6) \\
                                                            & 22.6(0.6) & 34.8(1.8) & 42.8(2.8) & 49.6(2.1) & 51.6(1.7) & 55.2(0.5) \\ \hline
  \end{tabular}
  \caption{LLC misses (\%) for \azu\ version and number of threads (T). The first row of each dataset corresponds to Shared-None version, second row to Shared-Bin and third row to Shared-All. Numbers in parenthesis are standard deviations in \%.}
  \label{tab:dstable}
\end{table}
\normalsize

\begin{figure}[htp]
  \centering
  \includegraphics[scale=.9,bb=62 91 299 340,clip]{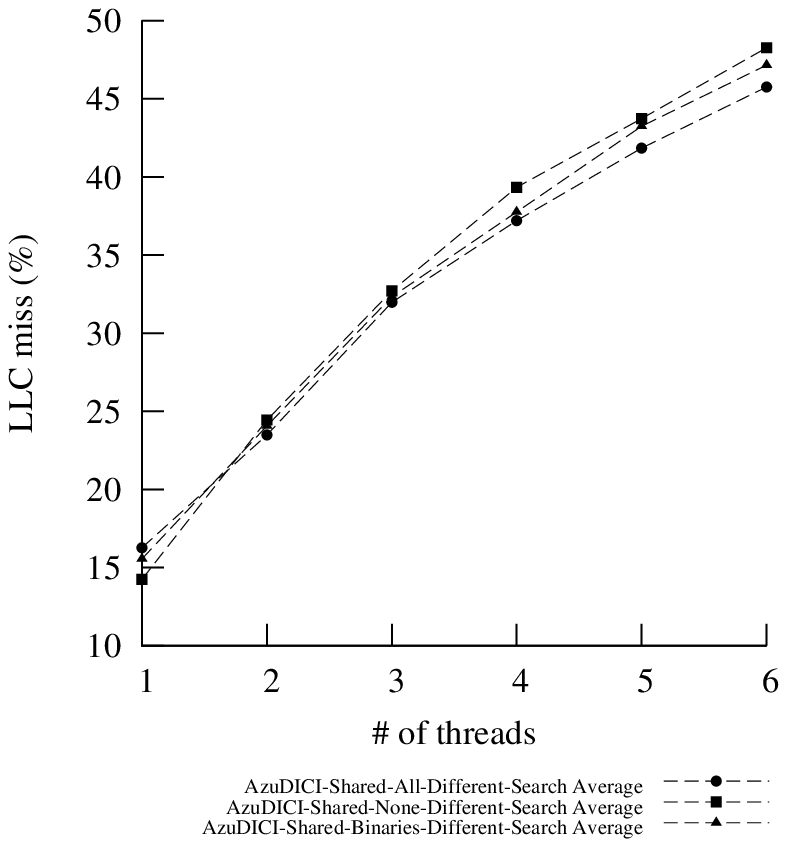}
  \caption{Average cache misses for AZUDici version and number of
    threads}
  \label{fig:dscachemisses}
\end{figure}

There are a few things to notice about Table \ref{tab:dstable}. First,
there are overall fewer cache misses in the shared-all setting, with
the exception of file {\tt grid-...-3.035-NOTKNOWN}. This file is
particular in that there is almost no cache contention. This may be
because the percentage of binary clauses (98\%) accounts for
practically all clauses. Due to the special data structures used for
propagation with binary clauses, the propagation computation does not
incur in noticeable cache penalties. It is also interesting to notice
that the difference between shared-binary and shared-all when compared
to shared-none (see Figure \ref{fig:dscachemisses}) is not as large as
in the previous section (see Figure \ref{fig:sscachemisses}, and Table
\ref{tab:sscachemisses} for details).
  
From these data, we may conclude that physically sharing the whole
clause database does not seem to significantly improve the cache
performance of the portfolio-based pSAT solvers.


\section{Conclusions and Future Work}
\label{sec:conclusions}

We showed that the impact of threads accessing the shared caches is
very significant and it negatively impacts the scalability of
portfolio-based pSAT solvers. We believe this is an important
topic for further advancing knowledge on parallel SAT solvers.

When it comes to whether physically sharing the clause database among
threads is advantageous, we may conclude that it is not yet clear
whether sharing the clause database in the way we have proposed has a
significant effect on running time of the solver. On the one hand, it
should be clear that physically sharing data (in general) should be
beneficial for parallel programs (if for no other reason, just to save
space). However, to implement solvers that physically share the clause
database is non-trivial, and prone to increasing complexity and
running time. Interestingly, the problems are not related, as is usual
with these kinds of systems, to, for instance, synchronization among the
threads, but rather stem from the BCP mechanism. This is so because
each thread must keep track of its own watches (which, as far as is
known, cannot be shared in a portfolio approach).

The relevance of cache efficient algorithms in CDCL SAT-solvers
performance has been known since at least 1993, with Zhang and Malik's
paper \cite{ZhangMalik2003SAT}. In this work we intend to update and
measure such influence in portfolio-based parallel CDCL SAT-solvers.
The design of \azu\ shared-all version is quite similar to that of
the {\tt (pa)MiraXT} solver \cite{paMiraXT}. To the best of our
knowledge, the first portfolio-based pSAT solver that physically shares
the clause database is {\tt SArTagnan} \cite{Sartagnan}. An in-depth survey
of parallel CDCL SAT-solvers can be found in \cite{survey-psolvers}.

As future work we also plan to continue the development of AzuDICI,
incorporating in it further enhancements like formula preprocessing
and variable elimination among others, so as to make it a competitive
parallel SAT solver. Regarding the Cache performance, we also plan to
implement our solver with compact data structures so that more
information can fit in the cache.

\section*{Acknowledgments}

We would like to thank several people and institutions for their help
in producing this paper, either in time, resources or both. First, we
would like to thank the Barcelogic group for sharing the code of a
past Barcelogic SAT-solver and for their support while writing
\azu. Likewise, we'd like to thank Armin Biere for opensourcing his
\pling\ code, and Enric Rodriguez-Carbonell, Technical University of
Catalunya for (conscientiously!) proofreading an earlier draft of this
paper. Finally, we would like to thank Paul Steinberg at Intel, and
the staff at the Intel ManyCore Testing Lab, where we carried out the
experiments. Roberto As\'in acknowledges the partial support
from Fondecyt through Project No. 11121220

\bibliographystyle{alpha} \bibliography{bibfile}

\end{document}